\documentclass[12pt,a4paper]{article}
\usepackage[latin1] {inputenc}
\usepackage[T1]{fontenc}
\usepackage{amsmath}
\usepackage{amsfonts}
\usepackage{amssymb}
\usepackage{latexsym}
\usepackage{pstricks}
\usepackage{pst-all}
\usepackage{url}
\usepackage[english]{babel}
\usepackage{latexsym}

\newcommand{\be}{\begin{equation}}
\newcommand{\ee}{\end{equation}}
\newcommand{\ba}{\begin{eqnarray}}
\newcommand{\ea}{\end{eqnarray}}

\newcommand{\dsl}{\kern.06em \hbox{ \raise.15ex \hbox{$/$}
\kern-.56em\hbox{$\partial$}}}

\newcommand{\eeq}{\end{equation}}
\newcommand{\eeqarr}{\end{eqnarray}}
\newcommand{\ZZ}{{\rm \kern 0.275em Z \kern -0.92em Z}\;}

\title{Noncommutative real scalar field theory in $2+1$ dimensions at finite
temperature}
\author{C.~D.~Fosco$^{a}$
and
G. A.~Silva$^{b}$\\
  {\normalsize\it $^a$Centro At\'omico Bariloche and Instituto Balseiro}\\
  {\normalsize\it Comisi\'on Nacional de Energ\'\i a At\'omica}
  \\
  {\normalsize\it R8402AGP Bariloche, Argentina.}\\
  ~
  {\normalsize\it $^b$IFLP/CONICET-Departamento de F\'\i sica}\\
  {\normalsize\it Universidad Nacional de La Plata}\\
  {\normalsize\it C.C. 67, 1900 La Plata, Argentina.}}
\begin{document}
\date{}
\maketitle
\begin{abstract}
\noindent
We study thermal effects for a noncommutative real scalar
field in $2+1$ dimensions including a Grosse-Wulkenhaar term.  Using
a perturbative expansion for the free energy, we deduce some general
properties of the corresponding contributions, in the thermodynamic
limit. We show that the model can be consistently interpreted as
defined on a finite volume, which is naturally determined by the
noncommutativity scale.
\end{abstract}
\section{Introduction}\label{sec:intro}

The class of Noncommutative Quantum Field Theories (NCQFT's)
obtained by a Moyal deformation of the usual (pointwise) product of
functions, has been extensively studied in recent
years~\cite{Douglas:2001ba,Sz}. One of the main reasons for
the renewed interest in this subject may be found in its relevance
to the dynamics of open string theories in non-vanishing constant
antisymmetric NS backgrounds, a set up that leads to noncommutativity
between the  string endpoint coordinates~\cite{SW}.

From a quite different standpoint, this topic has also attracted
attention because NCQFT's seem to be very good candidates for an
effective description of the Quantum Hall Effect (QHE)~\cite{Suss}.
Indeed, the existence of a strong magnetic field normal to an
(essentially) two-dimensional sample, paves the way to the use of
the Peierls substitution, whereby the two spatial coordinates
corresponding to each charged particle become noncommuting operators
\cite{Duval:2000xr}.

In these NCQFT's one usually considers quantum fields endowed with a
non-commutative (Moyal) $\star$ product which, for two functions $f(x)$ and
$g(x)$ ($x \in {\mathbb R}^{(d)}$), may be defined as follows:
\begin{equation}\label{eq:defmoyal}
\big( f \star g \big)(x) \;\equiv\;
\Big[ e^{ \frac{i}{2} \theta_{\mu\nu} \partial^x_\mu \partial^y_\nu} f(x)
\, g(y) \Big]_{y \to x} \;.
\end{equation}
Here, $\theta_{\mu\nu}$ is a constant antisymmetric tensor, and $\mu,\,\nu$
run over the spacetime indices. Since we shall be concerned with the case
of \mbox{$2+1$-dimensional}, i.e., `planar' theories, the
$\theta_{\mu\nu}$ tensor does necessarily have a zero mode. We shall assume
that this zero mode corresponds to the time-like ($\mu=0$) direction,
having in mind instances where the noncommutativity is, indeed, due to a
physical magnetic field, as in the QHE.  The $\theta_{\mu\nu}$ tensor will
thus verify $\theta_{0\mu} = 0$ and the time-like coordinate $x_0$ behaves,
to all effects, as a commuting object\footnote{This is particularly
relevant to thermal field theory since the temperature is a commuting
object so must therefore be the imaginary time $\tau$.}.

In this paper, we shall be concerned with the calculation of thermal
effects in the NCQFT of a real scalar field. Many interesting results have
recently been obtained by considering noncommutative systems obtained by a
Moyal deformation of a standard QFT, at a finite
temperature~\cite{wh,nct1,nct2,nct3}.  However, we want to consider here the
case of a model equipped with a Grosse-Wulkenhaar (GW)
term~\cite{Grosse:2003nw,Grosse:2004yu} (see also \cite{R}).  The reason
for considering this kind of theory, rather than standard ones is, in our
context, twofold.  Firstly, from a fundamental QFT point
of view (not necessarily at a finite temperature), we should do that in
order to solve the non-decoupling of UV and IR fluctuations that
unavoidably manifests in the absence of such a  term (a phenomenon usually
known as `IR-UV mixing'~\cite{MSV}). As a by product, one finds that a
NCQFT without a GW-term is, moreover, `anomalous' under the Langmann-Szabo
duality~\cite{LS}.

Secondly, and this is particularly relevant for the case at hand, one may
also want to include the GW-term because it effectively confines the system
to a finite volume, which is controlled by the strength of that term.
This situation is, indeed, of physical interest when one wants to consider
NCQFT at a finite temperature, in particular for the calculation of the free
energy. There, one has to assume that the theory has been defined on a
finite volume, and that volume tends to infinity at the end, when one takes
the thermodynamic limit. This is not an issue in the commutative case,
where one can simply use a box and impose periodic boundary conditions for
the fields. The situation is not so simple in the noncommutative case, however,
since one should have to face the problem of imposing boundary conditions
for the NCQFT defined in a finite volume, with the related technical
difficulties.

The GW-term introduces, for the case at hand, a kind of external
harmonic potential. It should be reminded that noncommutative
theories including a coupling to an external (usually magnetic)
field do have interesting properties, even in the context of
noncommutative quantum mechanics~\cite{Nair:2000ii,Bellucci:2001xp}.
Indeed, for a (noncommutative) charged particle an interesting
distinction between two different phases naturally emerges,
depending on the ratio between the external magnetic field strength
$B$ and the noncommutativity parameter $\theta$.  Those phases,
corresponding to Hamiltonians having  essentially different spectra,
are separated by a critical point  determined by a relation $\theta
B = \kappa$, where $\kappa$ is a numerical constant of order $1$,
whose precise value depends on the units and conventions adopted. At
that critical point the system becomes exactly solvable~\cite{LSZ}
(even in the presence of an external potential~\cite{Bellucci:2001xp}) and there
is, moreover, an effective `dimensional reduction'
\cite{Nair:2000ii}.

For the real scalar field we shall deal with in this work, there is
also an interesting interplay between the strength of the GW-term
$\Omega^2$ and the noncommutativity parameter $\theta$, although the
symmetry properties are quite different to the ones in the charged
field case.

In this paper we study the perturbative calculation of the free energy for
that model, focussing on the general properties of the first few terms,
providing explicit results whenever possible. We do that mostly
for the self-dual case ($\Omega=1$), and  discuss the relation between them
with the $\Omega \neq 1$ case.  The article is organized as follows: in
section~\ref{sec:free}, we present the perturbative calculation of the free
energy for a real scalar field in $2+1$ dimensions with a Grosse-Wulkenhaar
term.  In section~\ref{sec:concl}, we present our conclusions.
\section{Perturbative calculation of the free energy}\label{sec:free}
\subsection{General considerations}
Let us briefly review here the usual approach to the calculation of the
free energy, in the path-integral (imaginary-time)
context~\cite{Kapusta:2006pm}, to apply it afterwards to the real scalar
field case.

We shall start from the partition function ${\mathcal Z}$, depending
on \mbox{$\beta \equiv T^{-1}$} and, eventually (i.e., when there is
a non-trivial internal conserved charge)  on a chemical potential.
The functional integral representation of $\cal Z$ has the form
\begin{equation}\label{eq:defzgen}
{\mathcal Z} \;=\; \int {\mathcal D} \mu \; e^{- S } \;,
\end{equation}
where ${\mathcal D} \mu$ denotes the functional integration measure
corresponding to the space of fields being considered, which
requires periodic or antiperiodic conditions for the imaginary-time
coordinate \mbox{$x_0 \equiv \tau$}, according to the fields being
bosonic or fermionic, respectively.

In (\ref{eq:defzgen}), $S$ denotes the Euclidean action for a finite
imaginary-time interval, namely, $\tau \in [0,\beta]$. The actions that we
shall consider may be naturally decomposed as follows:
\begin{equation}\label{eq:sdec}
S \;=\; S_0 \,+\, S_I \;,
\end{equation}
where $S_0$ denotes the free, i.e., quadratic, part of the action,
and $S_I$ the interaction piece (at least cubic in the fields). Note
that the GW-term shall be thus included in $S_0$.

Based on the above decomposition for the action, one arrives to the
expression
\begin{equation}\label{eq:zspl}
{\mathcal Z} \;=\; {\mathcal Z}_0 \;\; \langle e^{- S_I} \rangle
\end{equation}
where
\begin{equation}\label{eq:defz0}
{\mathcal Z}_0 \;=\; \int {\mathcal D} \mu \; e^{- S_0} \;,
\end{equation}
and we have introduced the notation:
\begin{equation}\label{eq:defav0}
\langle \ldots  \rangle \;\equiv\; \frac{1}{\mathcal Z_0} \; \int
{\mathcal D} \mu \ldots  e^{- S_0}  \;,
\end{equation}
for Gaussian averages defined by the free action.

When constructing an expansion in powers of $S_I$, it turns out to
be simpler to consider the free energy ${\cal F} \equiv -
\frac{1}{\beta}\ln{\mathcal Z}$. Indeed, one easily finds that
\begin{equation}
\label{eq:lzdec}
{\cal F} \;=\; {\cal F}_0 \,+\, {\cal F}_I
\end{equation}
where
\begin{equation}
{\cal F}_0 = - \frac{1}{\beta} \ln {\mathcal Z_0}
\label{f0}
\end{equation}
and
\begin{equation}\label{eq:deflnzi}
{\cal F}_I \;\equiv\; - \frac{1}{\beta} \, \ln \, \langle e^{- S_I}
\rangle \;.
\end{equation}
Expanding ${\cal F}_I$ in powers of  $S_I$, one obtains
\begin{equation}
{\cal F}_I \;=\; {\cal F}_I^{(1)}\,+\, {\cal F}_I^{(2)}\,+\, {\cal
F}_I^{(3)} \,+\,\ldots
\end{equation}
where
\begin{equation}\label{eq:deffi1}
{\cal F}_I^{(1)} \,=\, \frac{1}{\beta}\, \langle S_I\rangle\;,
\end{equation}
\begin{equation}
{\cal F}_I^{(2)} \;=\; - \frac{1}{2!\beta}\, \Big\langle \big( S_I -
\langle S_I\rangle \big)^2 \Big\rangle \;,
\end{equation}
\begin{equation}
{\cal F}_I^{(3)} \;=\; \frac{1}{3!\beta}\,\left\langle \big( S_I
\,-\, \langle S_I\rangle \big)^3 \right\rangle \;, \ldots
\end{equation}

In the remaining parts of this section, we first define the model that we
shall study in detail, and afterwards we calculate the first few
terms in the above-defined expansion.
\subsection{The model}
The model is defined by an Euclidean action \mbox{$S=S_0+S_I$}, with
\begin{equation}\label{eq:defs0bos}
  S_0\;=\;\frac{1}{2}\int_0^\beta d\tau\int d^2x \, \big[\partial_\mu\phi \star
  \partial_\mu\phi\,+\, \Omega^2\, (\tilde x_i \phi) \star (\tilde x_i\phi)
  \,+\,m^2 \phi \star \phi  \big]
\end{equation}
where $\tilde x_i \equiv 2 (\theta^{-1})_{ij}\,x_j$, and
\begin{equation}\label{eq:defsibos}
  S_I \;=\; \frac{\lambda}{4 !} \int_0^\beta d\tau \int d^2x \, \phi \star \phi
  \star \phi \star \phi \;.
\end{equation}
We shall assume that $\theta >0$ and $\Omega \geq 0$ (without any lose of
generality).

The harmonic potential proportional to $\Omega^2$ in
(\ref{eq:defs0bos}) is the GW term. It has been shown that its confining
properties provide an infrared cutoff and, by the same token, tame the IR
problem due to IR/UV mixing~\cite{Grosse:2004yu}.

By using elementary properties of the Moyal product, the free action
$S_0$ may also be written in the equivalent way
\begin{eqnarray}\label{eq:s0boseq}
  S_0 =\frac{1}{2} \,\int_0^\beta d\tau\int d^2x \Big[\partial_\tau\phi \star
  \partial_\tau\phi\,+\,
  \big( \frac{1 + \Omega^2}{2} \big) \, \phi \star {\tilde x}_j \star {\tilde x}_j  \star \phi
  \,\nonumber\\- \big(\frac{1 - \Omega^2}{2}\big) \big({\tilde x}_j \star \phi
\star {\tilde x}_j \star \phi\big)
  +\,m^2 \phi \star \phi  \Big] \;.
\end{eqnarray}

\subsection{Zeroth order (`ideal gas')}
The zeroth-order term ${\cal Z}_0$ is obtained from the evaluation of a
Gaussian functional integral,
\begin{equation}
{\mathcal Z}_0 \;=\; \int {\mathcal D} \phi \; e^{- S_0 [ \phi ] } ~.
\label{z0}
\end{equation}
Following the usual procedure of QFT at finite temperature, we decompose the field
$\phi$, periodic in the time coordinate, in
terms of its Fourier components
\begin{equation}
\phi(x) \,=\, \phi(\tau,{\mathbf x}) \,=\, \beta^{-\frac{1}{2}} \;
\sum_{n=-\infty}^{+\infty} e^{i \omega_n \tau} \, \phi_n({\mathbf x}) \;,
\label{exp}
\end{equation}
with $\omega_n = \frac{2\pi n}{\beta}$. Since we are considering a real field,
$\phi^\dagger = \phi$, one has
\begin{equation}
\phi_0^\dagger({\mathbf x})\,=\, \phi_0 ({\mathbf x}) \;\;,\;\;\;
\phi_n^\dagger({\mathbf x})\,=\, \phi_{-n} ({\mathbf x}) \;\;,\;\;\forall n
\geq 1\;.
\end{equation}
In terms of these modes, the free action becomes a decoupled sum of ($d=2$)
actions, involving a real field $\phi_0$ and an infinite number of complex
fields $\phi_n$
\begin{equation}\label{eq:sbdec}
S_0[\phi] \;=\;S_0^{(0)}[\phi_0] \,+\, \sum_{n=1}^{\infty} \,
S_0^{(n)}[\phi_n^\dagger, \phi_n] \;,
\end{equation}
where
\begin{equation}
  S_0^{(0)}[\phi_0] \;=\;\frac{1}{2} \int d^2x
 \big[ \partial_j \phi_0 \star \partial_j\phi_0
  \,+\, \Omega^2 ({\tilde x}_j \phi_0) \star ({\tilde x}_j \phi_0)
  \,+\, m^2 \phi_0 \star \phi_0 \big]
  \label{s1}
\end{equation}
and
\begin{eqnarray}
  S_0^{(n)}[\phi_n^\dagger,\phi_n] &=& \int d^2x \big[ \partial_j \phi^\dagger_n \star \partial_j\phi_n
  \,+\, \Omega^2 ({\tilde x}_j \phi_n^\dagger) \star ({\tilde x}_j \phi_n)
  \nonumber\\
  &+& (m^2 + \omega^2_n)\,\phi_n^\dagger \star \phi_n \big]  \;\;, \;\;\; n \geq
  1\;.
  \label{s2}
\end{eqnarray}
For each one of these two-dimensional theories we use the matrix base \cite{gbv} to
expand the fields as
\begin{eqnarray}
\phi_n ({\mathbf x}) &=& \sum_{i,j=0}^\infty \, \phi_n^{(i,j)} \,
b^{(i,j)}({\mathbf x})  \nonumber\\
\phi_n^\dagger ({\mathbf x}) &=& \sum_{i,j=0}^\infty \, {\bar\phi}_n^{(i,j)} \,
b^{(j,i)}({\mathbf x}) \;.
\label{md}
\end{eqnarray}
The integration measure ${\mathcal D}\phi$ in (\ref{z0}) is defined in terms of the
Fourier components (\ref{exp}) as
\begin{equation}
{\mathcal D}\phi \;=\; {\mathcal D}\phi_0 \, \prod_{n=1}^\infty \, {\mathcal D}\phi_n^\dagger \,
{\mathcal D}\phi_n \;.
\end{equation}
The explicit form for them is
\begin{equation}
{\mathcal D}\phi_0 \;=\, \Big( \prod_i {\mathcal D}\phi_0^{(i,i)} \Big)  \;
\prod_{i<j}{\mathcal D}{\bar\phi}_0^{(i,j)} {\mathcal D}\phi_0^{(i,j)} \;,
\end{equation}
and
\begin{equation}
{\mathcal D}\phi_n^\dagger{\mathcal D}\phi_n \;=\; \prod_{ij}{\mathcal
D}{\bar\phi}_n^{(i,j)}
  {\mathcal D}\phi_n^{(i,j)} \;.
\end{equation}
In terms of the matrix base decomposition (\ref{md}), the actions (\ref{s1})-(\ref{s2})
take the form
$$
S_0^{(0)}[\phi_0] \;=\;(2 \pi \theta) \; \frac{1}{2} \,
\sum_{i_1,i_2;j_1,j_2} \, \phi_0^{(i_1,i_2)} {\mathcal G}_0^{(i_2,i_1;j_1,j_2)}
\phi_0^{(j_1,j_2)}
$$
\begin{equation}
= (2 \pi \theta) \; \frac{1}{2} \,
\sum_{i;j} \, \phi_0^{(i,i)} {\mathcal G}_0^{(i,i;j,j)}
\phi_0^{(j,j)}
\,+\,(2 \pi \theta) \, \sum_{i_1<i_2;j_1<j_2} \, {\bar\phi}_0^{(i_1,i_2)} {\mathcal
  G}_0^{(i_1,i_2;j_1,j_2)} \, \phi_0^{(j_1,j_2)}\label{s01}
\end{equation}
and
\begin{equation}
S_0^{(n)}[\phi_n^\dagger,\phi_n] \,=\, 2 \pi \theta \;
\sum_{i_1,i_2;j_1,j_2} \, {\bar\phi}_n^{(i_1,i_2)} {\mathcal G}_n^{(i_1,i_2;j_1,j_2)}
\phi_n^{(j_1,j_2)} \;\;,\;\;\; \forall n \geq 1\;,\label{s02}
\end{equation}
where\footnote{Note that our convention for the kernel differs with the
one used in~\cite{Grosse:2003nw,Grosse:2004yu} in a transposition of the
first pair of indices.}
\begin{eqnarray}\label{eq:defg}
{\mathcal G}_n^{(i_1,i_2;j_1,j_2)} &=& \big[ m^2 + \omega_n^2 + \mu^2 (i_1 + i_2 + 1) \big] \delta_{i_1j_1}
\delta_{i_2j_2} \nonumber\\
&-& \mu^2 \sqrt{\omega} \sqrt{ (i_1+1)  (i_2+1)} \,
\delta_{i_1+1,j_1} \delta_{i_2+1,j_2}
\nonumber\\
&-& \mu^2 \sqrt{\omega} \sqrt{ i_1  i_2} \, \delta_{i_1-1,j_1}
\delta_{i_2-1,j_2} \;,
\end{eqnarray}
with
\begin{equation}\label{eq:defmuo}
\mu^2 \;=\; 2 \frac{(1 + \Omega^2)}{\theta} \;\,,\;\;\;
\sqrt{\omega} \;=\; \frac{1 - \Omega^2}{1 + \Omega^2} \;.
\end{equation}
The integrals over the different Fourier modes decouple,
\begin{equation}
{\mathcal Z}_0 \;=\; \prod_{n=0}^\infty {\mathcal Z}_0^{(n)} \;,
\end{equation}
where
\begin{equation}\label{eq:zbos00}
{\mathcal Z}_0^{(0)}\;=\; \int {\mathcal D} \phi_0 \;
e^{-S_0^{(0)}[\phi_0]}
\end{equation}
and
\begin{equation}\label{eq:zbos0n}
{\mathcal Z}_0^{(n)}\;=\; \int {\mathcal D} {\phi}_n^\dagger \, {\mathcal D} \phi_n\,
e^{-S_0^{(n)}[\phi_n^\dagger,\phi_n]}\;.
\end{equation}
A careful use of the matrix base decomposition shows that the following
expressions hold true\footnote{Although we are taking the logarithm of
a dimensional quantity in (\ref{f}), a precise meaning can be given to the formula
using the analytic regularization technique \cite{Kleinert}.}
\begin{equation}
\ln {\mathcal Z}_0^{(0)} \;=\; -\frac{1}{2} \,{\rm Tr} \ln  {\mathcal G}_0
\end{equation}
and
\begin{equation}
\ln {\mathcal Z}_0^{(n)} \;=\; - \,{\rm Tr}\ln {\mathcal G}_n\;.
\end{equation}
In this last expressions the trace operation $\rm Tr$ should be understood as taken
over the whole set of matrix base elements $\{b^{(i,j)}\}$.
Thus,
\begin{eqnarray}
  \ln {\mathcal Z}_0 &=& -\frac{1}{2} \,{\rm Tr}\ln {\mathcal G}_0  -
  \sum_{n=1}^\infty {\rm Tr}\ln {\mathcal G}_n \nonumber\\
  &=& - \frac{1}{2} \sum_{n=-\infty}^\infty {\rm Tr}\ln {\mathcal G}_n \;.
  \label{f}
\end{eqnarray}
The sum over $n$ can of course be calculated (see for
example \cite{Kleinert}),
so that we may write the corresponding contribution to the free energy as
follows:
\begin{equation}\label{eq:k}
{\cal F}_0\;=\; \frac{1}{2} {\rm Tr} \sqrt{H} \,
+\, \beta^{-1} \, {\rm Tr} \ln\big( 1 - e^{-\beta \sqrt{H}}\big)
\end{equation}
where  an operator $H$ is introduced, such that its matrix elements
in the matrix basis $\{b^{(i,j)}\}$ are
\begin{eqnarray}\label{eq:defh}
H^{(i_1,i_2;j_1,j_2)}&\equiv&\int d^2x~ [b^{(i_1,i_2)}({\mathbf x})]^\dagger
H({\mathbf x})\,b^{(j_1,j_2)}({\mathbf x})\nonumber\\
 &=& \big( m^2 + \mu^2 (i_1 + i_2 + 1) \big) \delta_{i_1j_1}
\delta_{i_2j_2} \nonumber\\
&&- \mu^2 \sqrt{\omega} \sqrt{(i_1+1)(i_2+1)} \delta_{i_1+1,j_1}
\delta_{i_2+1,j_2}
\nonumber\\
&&- \mu^2 \sqrt\omega \sqrt{i_1 i_2}\, \delta_{i_1-1,j_1}
\delta_{i_2-1,j_2} \;.
\end{eqnarray}
In what follows, the first ($\beta$-independent) term in (\ref{eq:k}), which corresponds to the
vacuum energy part, shall be neglected since it does not contribute to the
thermodynamical properties of the system.

We proceed in the next paragraphs to evaluate ${\cal F}_0$ for
different values of the constant $\Omega^2$.  We first consider the
simplest $\Omega^2=1$ case, and then extend the result to the
general $\Omega^2\ne1$ situation by mapping the latter to the
former.

\subsubsection{The self-dual  $\Omega^2 =1$ case}
\label{sdc}

The $\Omega^2=1$ case becomes simple since $\omega $ given by
(\ref{eq:defmuo}) becomes zero and therefore the $H$ kernel is
diagonal. Explicitly,
\begin{equation}\label{eq:selfdg}
  H^{(i_1,i_2;j_1,j_2)} \;=\;\big(m^2 + \frac{4}{\theta} (i_1 + i_2 + 1)
  \big) \delta_{i_1j_1}\,\delta_{i_2j_2} \;.
\end{equation}
It's  eigenvalues $h_l$  are
\begin{equation}
h_l  =  \big(m^2 + \frac{4}{\theta} (l + 1) \big)  \;,\;\; l =
0,1,2,\ldots
\end{equation}
with a degeneracy equal to $l+1$ (the number of different ways to
obtain an integer $l$ by adding two non-negative integers $i_1$ and
$i_2$). The degeneracy in the energy for the free theory can be
seen to have its origin
in the existence of a dynamical $SU(2)$ symmetry for the two
dimensional isotropic oscillator (more on this below, see sect. \ref{ring}). The free energy of the self-dual
model is then
\begin{equation}
\Big[ {\cal F}_0 \Big]_{\Omega^2=1} \;=\; \beta^{-1} \,
\sum_{l=1}^\infty \, l \, \ln \Big( 1 \,-\, e^{-\beta \sqrt{m^2 +
\frac{4}{\theta} l}} \Big) \;.
\label{freesd}
\end{equation}

\subsubsection{The general case}
The general case corresponding to an arbitrary value of $\Omega^2$
can also be dealt with exactly.  We first note that the matrix
elements of $H$ may be regarded as the ones corresponding to an
Hermitian operator constructed out of two independent sets of
harmonic oscillator annihilation and creation operators $a_\alpha$,
$a_\alpha^\dagger$  ($\alpha=1,2$), as follows:
\begin{equation}\label{eq:defhop}
 H \;=\; m^2 + \mu^2 \big(a_1^\dagger a_1 \,+\, a_2^\dagger a_2 \,+\,1\big)
 \,-\,
\mu^2 \, \sqrt{\omega} \big( a_1 a_2 \,+\, a_1^\dagger a_2^\dagger \big) \;.
\end{equation}
The form of (\ref{eq:defhop}) will be simplified by performing a
Bogoliubov transformation. We first introduce a
two-component vector ${\mathbf a}$:
\begin{equation}
{\mathbf a} \;=\; \left(
\begin{array}{c}
a_1 \\
a_2^\dagger
\end{array}\right),
\end{equation}
from which a new two-component operator  ${\mathbf a}(\alpha)$ is
obtained by performing the (unitary) Bogoliubov transformation
\begin{equation}
{\mathbf a}(\alpha) \;=\; U^\dagger(\alpha) \, {\mathbf a} \, U(\alpha)
\end{equation}
with
\begin{equation}
U(\alpha) \;=\;e^{i \,\alpha\, G}
\end{equation}
and  the infinitesimal generator  $G$ given
by
\begin{equation}
G \;=\; i ( a_1 a_2 - a_2^\dagger a_1^\dagger ) \;.
\end{equation}
The   transformation $U(\alpha)$ maps the original operators $a_i$
to new ones $b_i$, such that
\begin{equation}
{\mathbf b} \equiv {\mathbf a}(\alpha) \,=\,
\left(
\begin{array}{cc}
\cosh\alpha & \sinh \alpha \\
\sinh\alpha & \cosh \alpha
\end{array}
\right) {\mathbf a} \;. \label{bogoliub}
\end{equation}
We shall fix  the hyperbolic angle $\alpha$ by requiring the
transformed operator $H(\alpha) \equiv U^\dagger(\alpha) \, H \,
U(\alpha)$, to be diagonal in terms of the new operators $b_i$. This
amounts to the equation
\begin{equation}
\tanh (2\alpha) \;=\; \frac{\Omega^2-1}{\Omega^2 + 1} \;,
\end{equation}
to be satisfied.
In terms of the new operators, the rotated Hamiltonian
$H(\alpha)$ adopts the form
\begin{eqnarray}
  H(\alpha) &=& m^2\,+\,\frac{\mu^2}{\cosh 2\alpha} \,
\big(b_1^\dagger b_1 \,+\, b_2^\dagger b_2 \,+\,1\big) \nonumber\\
  &=& m^2 \,+\, \frac{4 \Omega}{\theta} \,
\big(b_1^\dagger b_1 \,+\, b_2^\dagger b_2 \,+\,1\big) \;.
\label{diag}
\end{eqnarray}

It is then immediate to obtain the free energy in the general case from the
one corresponding to the self-dual case by making the substitution
$\frac4\theta\to\frac{4\Omega}\theta$ in (\ref{freesd}),
\begin{equation}
{\cal F}_0\;=\; \beta^{-1} \, \sum_{l=1}^\infty \, l \, \ln \Big( 1
\,-\, e^{-\beta \sqrt{m^2 + \frac{4 \Omega}{\theta} l}} \Big) \;.
\label{generalF}
\end{equation}

It is instructive to consider the small $\Omega$ limit of the
expression above, since we expect it to be related to the
corresponding free energy in the absence of the harmonic
$\Omega^2$-term. Besides, since this term plays the role of a
confining potential, it naturally defines an effective
physical volume $V$ of
order $\theta/(4 \Omega)$. To simplify matters, we
assume  $m=0$. For $\Omega \ll 1$, we approximate the sum in
(\ref{generalF}) by an integral, by making an elementary change of
variables, one obtains,
\begin{equation}
{\cal F}_0\;\simeq \; T^3 \, V \, \frac{\theta T^2}{4\Omega} \,
\int_0^\infty dy\,  y \, \ln \big( 1 \,-\, e^{-\sqrt y} \big) \;.
\end{equation}
The free energy density $f_0$ is therefore
\begin{equation}
f_0\;\simeq \; T^3 \,\frac{\theta T^2}{4\Omega} \, \int_0^\infty
dy\, y \, \ln \big( 1 \,-\, e^{-\sqrt y} \big) \;. \label{f0NC}
\end{equation}
This should be compared with the free energy density for a free
commutative scalar field in a box (which coincides with the result
for a free noncommutative theory in an infinite volume with no
harmonic term)
\begin{equation}
\Big[ f_0\Big]_{comm}\;=\; T^3 \, \frac{1}{2 \pi}
\int_0^\infty dy \,  y \, \ln \big( 1 \,-\, e^{-y} \big) \;.
\label{f0C}
\end{equation}

An important qualitative difference between the
results (\ref{f0NC}) and (\ref{f0C}) can be noted
due to the appearance of a (dimensionless) factor
\mbox{$\frac{\pi\theta T^2}{2\Omega}$}.

The $\Omega \to 0$ limit depends then on whether one takes it before
or after evaluating the free energy. Indeed, a free NCQFT without
the GW term ($\Omega = 0$) yields (\ref{f0C}), which coincides with
the result for a free commutative QFT. On the other hand, we see
that taking the $\Omega\to0$ limit after the evaluation of the free
energy yields instead a divergent result and does not match the
$\Omega=0$ result. This behavior is reminiscent to the well known
subtlety when taking the commutative $\theta\to0$ limit of
non-commutative theories \cite{MSV}.

We conclude this subsection by noting that the transformation that
diagonalices $H$ is of the type considered when dealing with
`two-mode squeezed states', in a quite different
context~\cite{thermo}. We can however, take advantage of the
equivalent form of the transformation in order to extract
conclusions for the case at hand.

As an example, the ground state of the model for arbitrary $\Omega$
may be obtained as follows: since the transformed operator
$H(\alpha)$  in (\ref{diag}) is diagonal in   the $b_i$  basis, we
may read off its `ground state' $|0(\alpha)\rangle$ as the one
annihilated by the $b_i$ operators. Taking advantage of the relation
(\ref{bogoliub}) it is therefore possible to write an explicit
relation between the ground state for arbitrary $\Omega$,
$|0\rangle_{\Omega}\equiv|0(\alpha) \rangle$, and the ground state
$|0\rangle_{\Omega=1}\equiv|0\rangle$ of the self-dual  Hamiltonian
(\ref{eq:selfdg}). The relation between them can be shown to be
given by \cite{thermo},
\begin{equation}
|0(\alpha) \rangle \;=\; \frac{1}{\cosh\alpha}\, e^{-\tanh \alpha \,
a_1^\dagger a_2^\dagger} \, |0\rangle \;.
\end{equation}
Or, in terms of the model parameters,
\begin{equation}
|0\rangle_\Omega \;=\; \frac{2\sqrt\Omega}{1+ \Omega}\, e^{-|\frac{1
- \Omega}{1 + \Omega}| \, a_1^\dagger a_2^\dagger} \,
|0\rangle_{\Omega = 1} \;.
\end{equation}
This  is an example of a two-mode squeezed state
(see~\cite{thermo}).

\subsection{First-order term}

We proceed now to evaluate the first-order contribution to the free
energy, ${\cal F}_I^{(1)}$. The expression for (\ref{eq:deffi1}) in
terms of the (Matsubara) Fourier components  for the field $\phi$ is
\begin{equation}
{\cal F}_I^{(1)} = \frac{2 \pi \lambda \theta}{4! \beta^2} \,
\sum_{n_1,\ldots,n_4=-\infty}^{+\infty}\!\!
\sum_{~~i_1,\ldots,i_4=0}^{+\infty} \; \delta_{\sum_i n_i=0} \,
     \langle
    \phi_{n_1}^{(i_1, i_2)} \phi_{n_2}^{(i_2, i_3)} \phi_{n_3}^{(i_3,i_4)}
\phi_{n_4}^{(i_4,i_1)}
     \rangle \;.
     \label{fi}
\end{equation}
Moreover, since the averages above are defined by a quadratic
weight, an application of Wick's theorem yields,
\begin{eqnarray}
\langle \phi_{n_1}^{(i_1,i_2)} \phi_{n_2}^{(i_2,i_3)}
\phi_{n_3}^{(i_3,i_4)} \phi_{n_4}^{(i_4,i_1)} \rangle &=&
\langle \phi_{n_1}^{(i_1,i_2)} \phi_{n_2}^{(i_2,i_3)} \rangle \;
\langle \phi_{n_3}^{(i_3,i_4)} \phi_{n_4}^{(i_4,i_1)} \rangle  \nonumber\\
&+& \langle \phi_{n_1}^{(i_1,i_2)} \phi_{n_3}^{(i_3,i_4)} \rangle \;
\langle \phi_{n_2}^{(i_2,i_3)} \phi_{n_4}^{(i_4,i_1)} \rangle \nonumber\\
&+& \langle \phi_{n_1}^{(i_1,i_2)} \phi_{n_4}^{(i_4,i_1)} \rangle \;
\langle \phi_{n_2}^{(i_2,i_3)} \phi_{n_3}^{(i_3,i_4)} \rangle \;.
\end{eqnarray}
Since the quadratic action $S_0$ only mixes Fourier components such
that their Matsubara indices add up to zero, we have,
$$
{\cal F}_I^{(1)} \;=\;\frac{2 \pi \lambda \theta}{4! \beta^2} \,
\sum_{n_1,n_2}  \sum_{i_1,\ldots,i_4} \Big( \langle
\phi_{-n_1}^{(i_1,i_2)} \phi_{n_1}^{(i_2,i_3)} \rangle \; \langle
\phi_{-n_2}^{(i_3,i_4)} \phi_{n_2}^{(i_4,i_1)} \rangle
$$
\begin{equation}
+\;\langle \phi_{-n_1}^{(i_1,i_2)} \phi_{n_1}^{(i_3,i_4)}\rangle \;
\langle \phi_{-n_2}^{(i_2,i_3)} \phi_{n_2}^{(i_4,i_1)} \rangle
+\langle \phi_{-n_1}^{(i_1,i_2)} \phi_{n_1}^{(i_4,i_1)} \rangle \;
\langle \phi_{-n_2}^{(i_2,i_3)} \phi_{n_2}^{(i_3,i_4)} \rangle \Big)
\;.
\end{equation}
Relabeling indices and combining identical terms we end up with,
\begin{eqnarray}
{\cal F}_I^{(1)} &=&\frac{2 \pi \lambda \theta}{4! \beta^2} \,
\sum_{n_1,n_2} \sum_{i_1,...,i_4} \Big( 2 \, \langle
\phi_{-n_1}^{(i_1,i_2)} \phi_{n_1}^{(i_2,i_3)} \rangle \;
\langle \phi_{-n_2}^{(i_3,i_4)} \phi_{n_2}^{(i_4,i_1)} \rangle \nonumber\\
&+& \, \langle \phi_{-n_1}^{(i_1,i_2)} \phi_{n_1}^{(i_3,i_4)}\rangle
\; \langle \phi_{-n_2}^{(i_2,i_3)} \phi_{n_2}^{(i_4,i_1)} \rangle
\Big) \;. \label{f1}
\end{eqnarray}
The first term can be interpreted as a planar graph in standard
double line notation (see \cite{MSV}), while the second one can be
seen to be non-planar. In the expressions above, the two point free
correlation function is determined from the quadratic action
(\ref{eq:defs0bos}), and its form strongly depends on the value of
$\Omega$.

As we did for the zero-order term, we perform in the following
section the explicit calculation for the self-dual point. We shall
then comment on the general case computation.

\subsubsection{Self-dual case}

As discussed in section \ref{sdc} the $\Omega =1$ case is
particularly simple since  the quadratic part of the action is
diagonal. We now proceed to compute the first-order term (\ref{f1})
taking into account that the two point correlation function adopts,
for the self-dual case, the quite simple form,
\begin{eqnarray}
\langle \phi_{-n}^{(i_1,i_2)} \phi_n^{(j_1,j_2)} \rangle &=&
\langle {\bar\phi}_n^{(i_2,i_1)} \phi_n^{(j_1,j_2)} \rangle  \nonumber\\
&=& \frac{1}{2\pi\theta} \,
\big( m^2 \, +\, \omega_n^2 \,+\, \frac{4}{\theta} (i_1 + i_2 + 1)
\big)^{-1}\; \delta_{i_1j_2} \, \delta_{i_2j_1} \;.
\label{propagator}
\end{eqnarray}
We separate the planar and non-planar contributions as,
\begin{equation}
{\mathcal F}_I^{(1)} \;=\; P\,+\,Q \label{f1corr}
\end{equation}
where
\begin{eqnarray}
P &=& \frac{\lambda}{4! \beta^2 \pi\theta} \sum_{n_1,n_2}
\sum_{i_1,i_2,i} \Big[ \big( m^2 \, +\, \omega_{n_1}^2 \,+\,
\frac{4}{\theta} (i_1+ i +1) \big)^{-1}\nonumber\\
&& \times~ \big( m^2 \, +\, \omega_{n_2}^2 \,+\, \frac{4}{\theta}
(i_2 + i + 1) \big)^{-1} \Big] \;, \label{planar}
\end{eqnarray}
and
\begin{eqnarray}
Q &=& \frac{\lambda}{4!\,2\pi\theta\, \beta^2 } \sum_{n_1,n_2}
\sum_{i} \Big[ \big( m^2 \, +\, \omega_{n_1}^2 \,+\,
\frac{4}{\theta} (2 i + 1)
\big)^{-1} \nonumber\\
&\times& \big( m^2 \, +\, \omega_{n_2}^2 \,+\, \frac{4}{\theta} ( 2
i + 1) \big)^{-1} \Big] \;. \label{nonplanar}
\end{eqnarray}
The $i$'s indices structure in the last two equations manifest the
expected worse UV behavior for the planar contribution
(\ref{planar}) as compared to the non-planar one (\ref{nonplanar})
\footnote{The number of sums in $i$ for any diagram can be seen to
be equal, when drawing it in double line notation, to the number of
independent loops.}.

It is useful for what follows to simplify the expressions
(\ref{planar})-(\ref{nonplanar}). We first note that the planar
contribution may be written as,
\begin{equation}
P \;=\; \frac{\lambda}{4! \pi\theta} \; \sum_{i=0}^{\infty} \big[
\,\mathcal{S}(i) \,\big]^2\;,
\label{p}
\end{equation}
where
\begin{equation}
\mathcal{S}(i) \;=\;\frac{1}{\beta} \, \sum_{n=-\infty}^{\infty}
\sum_{j=0}^\infty \big( m^2 \, +\, \omega_n^2 \,+\, \frac{4}{\theta}
(i + j + 1) \big)^{-1} \;. \label{s}
\end{equation}
For the non-planar term, we have instead,
\begin{equation}
Q \;=\;\frac{\lambda}{4! 2 \pi\theta} \sum_{i=0}^{\infty} \big[
\mathcal{T}(i) \big]^2\;,
\label{q}
\end{equation}
where
\begin{equation}
\mathcal{T}(i) \;=\;\frac{1}{\beta} \, \sum_{n=-\infty}^{\infty}
\big( m^2 \, +\, \omega_n^2 \,+\, \frac{4}{\theta} (2 i + 1)
\big)^{-1} \;. \label{t}
\end{equation}

It is evident that there are UV divergences lurking in the
expressions (\ref{p})-(\ref{q}), and we now deal with them. We
will show below that to first order in $\lambda$, as it happens in
the commutative case at finite temperature (see
\cite{Kapusta:2006pm}), only a mass counterterm is required to give
meaning to the free energy.

Let us begin computing the  two point function counterterm to first
order in $\lambda$, since it should be taken into account in the
calculation of the free energy we performed above. It is
straightforward to see that the divergent contribution to the
quadratic part of the effective action comes from a planar tadpole
diagram and takes the form,
\begin{equation}
\left.\phantom{\int}\Gamma_2[\phi]\right|_{div} \;=\;
\frac{\lambda}{3! (2\pi\theta)} {\mathcal S}_{0}(i\!_s) \int_0^\beta
d\tau \,\int d^2 x \,\phi \star \phi\;,
\label{div}
\end{equation}
where ${\mathcal S}_0(i)$ can be identified as the zero temperature
part of ${\mathcal S}(i)$, namely,
\begin{equation}
{\mathcal S}_0(i) \;=\; \int \frac{d\omega}{2\pi} \,
\sum_{k=0}^\infty \, \frac{1}{\omega^2 + m^2 + \frac{4}{\theta} ( k
+ i + 1)} \;,
\end{equation}
easily seen to be UV divergent for any arbitrary integer $i$
(interpreted in (\ref{div}) as the substraction point). Performing the integral over
$\omega$, one can write this last expression as,
\begin{equation}
{\mathcal S}_0(i) \;=\; \frac{\sqrt{\theta}}{4} \;
\zeta(\frac{1}{2}, \frac{ m^2 \theta }{4} + i + 1) \;,
\label{zeta}
\end{equation}
where now, $\zeta(s,a)$ should be understood as the analytical
continuation in $s$ of the generalized Riemann (Hurwitz) $\zeta$
function. The analytical regularization renders a finite result for
(\ref{zeta}),  the (infinite) ambiguity constant of which is fixed
by a renormalization condition for the mass parameter.

We therefore fix the mass counterterm to have the form\footnote{We fix
the substraction point $i\!_s$ in (\ref{div}) to be zero.},
\begin{equation}
 \delta m^2 \;=\; - \,
 \frac{\lambda}{3! \pi\theta}  {\mathcal S}_0(0) \;.
 \label{mct}
\end{equation}
This counterterm cancels the relevant divergences in $P$ when
the regularization is removed. Separating ${\mathcal S}(i)$ into its
zero temperature part ${\mathcal S}_0$ plus its thermal contribution
${\mathcal S}_T$ (which vanishes at $T=0$), we get,
\begin{eqnarray}
P &=& \frac{\lambda}{4! \pi\theta} \, \sum_{i=0}^{\infty}
\,\mathcal{S}_T (i)\, \mathcal{S}_T (i)
\nonumber\\
&+& \frac{ 2 \lambda}{4! \pi\theta} \, \sum_{i=0}^{\infty}
\,\mathcal{S}_0 (i)\, \mathcal{S}_T (i) \, \;+\;\frac{\lambda}{4!
\pi\theta} \, \sum_{i=0}^{\infty} \,\mathcal{S}_0 (i)\,
\mathcal{S}_0 (i) \;.
\end{eqnarray}
The last term will be ignored from now on, since we are not
interested in temperature-independent terms (which amount to vacuum
energy contributions).  The divergence in the second term is, on the
other hand, exactly canceled by the chosen mass counterterm
(\ref{mct}). Thus, the renormalized $P$ becomes,
\begin{equation}
P_{ren} \;=\; \frac{\lambda}{4! \pi\theta} \,
\sum_{i=0}^{\infty} \,\mathcal{S}_T (i) \mathcal{S}_T (i) \,
\,+\,\frac{ 2 \lambda}{4! \pi\theta} \,
\sum_{i=0}^{\infty} \,\tilde{\mathcal S}_0 (i) \mathcal{S}_T (i) \,
\end{equation}
where $\tilde{\mathcal S}_0 (i)\equiv \mathcal{S}_0 (i) -  \mathcal{S}_0
(0)$.

On the other hand, the divergent part of $Q$ gets  canceled by just
subtracting the  temperature independent (vacuum-energy)
contribution. Indeed, splitting ${\mathcal T}$ as we did for
${\mathcal S}$, we have,
\begin{equation}
Q_{ren} \;=\; \frac{\lambda}{4! 2 \pi\theta} \,
\sum_{i=0}^{\infty} \,\mathcal{T}_T (i) \mathcal{T}_T (i) \,
\;+\; \frac{ 2 \lambda}{4! 2 \pi\theta} \,
\sum_{i=0}^{\infty} \,{\mathcal T}_0 (i) \mathcal{T}_T (i)
\;,
\end{equation}
where (contrary to what happened in the calculation of $P_{ren}$) there is
no contribution to $Q$ from the counterterm (\ref{mct}), since its insertion
yields, to this order, only planar contributions.

Explicitly the function $\tilde{\mathcal S}_0(i)$ takes the form,
\ba \tilde{\mathcal S}_0(i)&=& \frac{\sqrt{\theta}}{4} \; \big[
\zeta(\frac{1}{2}, \frac{m^2 \theta}{4} + i + 1 ) \,-\,
\zeta(\frac{1}{2}, \frac{m^2 \theta}{4} +  1) \big]\nonumber\\
&=&
-\frac{\sqrt{\theta}}{4} \sum_{k=1}^i
\frac{1}{ \sqrt{k+\frac{ m^2\theta}{4}}} \;.
\ea
The temperature dependent piece, ${\mathcal S}_T(i)$ is, on the
other hand, given by
\begin{eqnarray}
{\mathcal S}_T(i) &=& \frac{\sqrt{\theta}}{2} \, \sum_{j=0}^\infty
\frac{1}{\sqrt{\frac{ m^2\theta}{4} + j + i +1}} \,
\frac{1}{e^{\frac{2\beta}{\sqrt{\theta}} \sqrt{\frac{ m^2\theta}{4} + j + i
+1} } - 1} \nonumber\\
&=&
\frac{\sqrt{\theta}}{2} \, \sum_{j=i}^\infty
\frac{1}{\sqrt{\frac{ m^2\theta}{4} + j +1}} \,
\frac{1}{e^{\frac{2\beta}{\sqrt{\theta}} \sqrt{\frac{ m^2\theta}{4} + j
+1}} - 1} \;.
\end{eqnarray}
For ${\mathcal T}(i)$ we obtain,
\begin{equation}
{\mathcal T}_0(i) \;=\;\frac{\sqrt{\theta}}{4} \,
\frac{1}{\sqrt{\frac{ m^2\theta}{4} + 2 i +1}}
\end{equation}
and
\begin{equation}
{\mathcal T}_T(i) \;=\;\frac{\sqrt{\theta}}{2} \,
\frac{1}{\sqrt{\frac{ m^2\theta}{4} + 2i +1}} \,
\frac{1}{e^{\frac{2\beta}{\sqrt{\theta}} \sqrt{\frac{ m^2\theta}{4} + 2i
+1} } - 1} \;.
\end{equation}
Inserting the previous expressions into $P_{ren}$ and $Q_{ren}$ yields the
first order contribution in $\lambda$  to the free energy
as a combination of multiple series, which cannot, in general, be
summed in closed form. However, for the massless case it is easy to see that
the first order correction will have the form
\begin{equation}
{\mathcal F}_I^{(1)} \;=\; \lambda \, f(\sqrt{\theta} T) \;.
\end{equation}
This is, $\lambda$ times  a function of the dimensionless combination involving
the noncommutativity parameter $\theta$ and the temperature $T$. In the limit
$\sqrt{\theta} T >> 1$, one can also see that the leading behavior of the free energy
is of the form $f(x) \sim x^2$, thus,
\begin{equation}
{\mathcal F}_I^{(1)} \;\sim\; \lambda \, \theta \, T^2 \;\;.
\end{equation}
This leading contribution comes only from the planar diagrams $P$, as
the non-planar ones $Q$ vanish in this limit.

It is worth noting that, in the massless limit, the corresponding
contribution for the commutative analogue of this model has a similar form,
\begin{equation}
\big[{\mathcal F}_I^{(1)} \big]_{comm}\;= \; \kappa \, \lambda \, L^2 \,  T^2 \;,
\end{equation}
where $L^2$ is the `volume' (i.e., area) of the system, and $\kappa$
is a numerical constant. The noncommutativity parameter $\theta$ appears again
as playing the role of the area of the system, at least when that parameter
is big in comparison with the temperature. This is our second
argument for interpreting $\frac{\theta}{4\Omega}$ (here $\Omega =
1$) as the `volumen' of the system.

\subsubsection{General case}

For general $\Omega$ the complicated form of the propagator renders
the calculation of first order computations quite involved. However,
when considering the free energy in the large $\theta$
(`thermodynamic') limit, an important simplification arises. Indeed,
in this limit only the planar diagrams contribute: the reason (as it happened
for the self-dual case) is that non-planar graphs, having less
independent summations, have a softer scaling behavior in the
thermodynamic limit, and therefore get suppressed. Since
only planar diagrams are relevant, one can then take advantage of
the Bogoliubov transformation of the free case, which allows one to
map the non self-dual case to the dual one, by a simple rescaling of
$\theta$ in the propagators. For the planar contribution (and only
for them) the unitary operators cancel out. Thus, in the
thermodynamic limit, the result for the free energy correction ${\cal F}^{(1)}_I$  only differ from a  $\frac{1}{\Omega}$
factor in comparison with the self-dual contribution (as it was the
case for the ideal gas term).

\subsection{Summation of the planar ring diagrams for the self-dual case}
\label{ring}

We conclude this section by considering the summation of the planar ring
diagrams (non planar are discarded, since we have in mind the
large-volume limit). We do this in order to exhibit some of the
peculiarities of the present model.

The second order term in the (renormalized)  effective action is given by,
\begin{eqnarray}
\big[\Gamma_2(\phi)\big]_{ren}&=& \frac{\lambda}{2 \times 3!} \,
\int_0^\beta d\tau \, \sum_{i,j=0}^\infty \,
\phi^{(j,i)}(\tau) \big[ \tilde{\mathcal S}_0(i) + {\mathcal S}_T(i)
\nonumber\\
&+& \tilde{\mathcal S}_0(j) + {\mathcal S}_T(j) \big]
\phi^{(i,j)}(\tau) \;,
\end{eqnarray}
where we have kept the same notations as in the previous subsection.

We note that, being this contribution diagonal in the matrix base, the
contribution corresponding to the summation of the ring diagrams can be
written straightforwardly. Indeed, it corresponds to the calculation of
the Gaussian functional integral that results from the inclusion of the
quadratic term in the effective action, and subtracting the (already
written) lower-order terms, to avoid double counting. The expression can
be put in the following way,
\begin{eqnarray}
{\mathcal F}_{ring} &=& \frac{1}{2 \beta}\, \sum_{n=-\infty}^\infty\sum_{i,j=0}^\infty
\ln \Big\{ \omega_n^2 \,+\, m^2 \,+\, \frac{4}{\theta} (i + j + 1)
\nonumber\\
&+& \frac{\lambda}{3! 2 \pi \theta} [ \tilde{\mathcal S}_0(i) + {\mathcal S}_T(i)
+ \tilde{\mathcal S}_0(j) + {\mathcal S}_T(j) ] \Big\} \nonumber\\
&-& {\mathcal F}_0 \;-\; P \;.
\end{eqnarray}
Note that, for $n=0$ and $m=0$, the would-be IR divergent contributions are
not only cured by the first order correction self-energy term, but also by the $\theta^{-1}$
factor always present whenever there is a finite volume, whose origin is the zero point energy
of the two dimensional oscillators.
The summation over the Matsubara frequencies can be performed leading to,
\begin{equation}
{\mathcal F}_{ring} \;=\; \beta^{-1}\, \sum_{i,j=0}^\infty
e^{- \beta \sqrt{ m^2 \,+\,{\mathcal E}_{i,j}} } \;-\;{\mathcal F}_0 \;-\; P \;.
\end{equation}
where
\begin{equation}
{\mathcal E}_{i,j} \;=\;  \frac{4}{\theta} (i + j + 1)
+\frac{\lambda}{3! 2 \pi \theta} [ \tilde{\mathcal S}_0(i) + {\mathcal S}_T(i)
+ \tilde{\mathcal S}_0(j) + {\mathcal S}_T(j) ] \;.
\end{equation}
We conclude by mentioning an important outcome of this expression:
since each element in the sum is no longer a function of $i+j$, the
degeneracy we had for the free case is lifted. This may be thought of as
due to the fact that, when including the $\phi^4_\star$ interaction,
the dynamical $SU(2)$ symmetry of the free theory cannot be preserved.

\newpage

\section{Conclusions}\label{sec:concl}

We have considered the perturbative calculation of the free energy
for a noncommutative real scalar field theory in $2+1$ dimensions in
the presence of a Grosse-Wulkenhaar term. We have first shown, at
the free (`ideal gas') level, that the free energy for the GW-model
has a qualitatively different temperature behavior when compared
to the known result obtained by assuming the noncommutative
theory to be  defined on an
infinite volume from the very beginning. The qualitative difference
being due to the appearance of the dimensionless parameter
$\frac{\theta T^2}{4\Omega}$ in (\ref{f0NC}). Moreover, for the known infinite
volume  ($\Omega=0$) case,
the free energy turns out to be $\theta$-independent since the
$\theta$-dependence can solely arise from boundary terms which
vanish for our choice of boundary conditions.

Of course, one might
have also considered the $\Omega=0$ case in a finite-volume
situation. That procedure should also produce, we believe, a non
trivial large-volume limit, due to the interplay between the
noncommutativity and boundary conditions.

Regarding the perturbative corrections, in spite of the difficulties
to obtain analytical results, some general properties clearly
emerge. Firstly, the perturbative computations and the harmonic
potential form of the GW-term suggest to interpret the volume of the
system to be given  by $V\sim\theta/4\Omega$. Secondly, in the
thermodynamic $\theta\to\infty$ limit,  only planar graphs yield a
non-vanishing contribution. Moreover, for the arbitrary $\Omega$
case, one can see that, again in the thermodynamic limit, the
contribution of the planar graphs, can be obtained from the
calculation of the ones for the self-dual case, by a redefinition of
the propagator, which essentially amounts to a rescaling of
$\theta$.

Finally, we have constructed a series that represent the sum of the
planar ring diagrams, showing how the GW-term moderates its IR
behavior. As an outcome, the calculation shows that the dynamical $SU(2)$
symmetry present in the free theory is not preserved when interactions are turned on.

The renormalization process that gives meaning to the perturbative
computations goes in complete analogy with the commutative case.
We have shown that, as in the commutative case,  at the first perturbative
order no new divergences appear at finite temperature, and the expressions
get regularized, if  the divergences at zero temperature where already
tamed. We should mention nevertheless, that contrary to the standard $\Omega=0$ case, the planar
tadpole contributions to the self energy depend on the external momentum.

\section{Acknowledgments}
G.A.S. would like to thank Glenn Barnich and Andy Gomberoff for email
correspondence. C.D.F. thanks CONICET and ANPCyT for financial support and
G.A.S.  acknowledges
support from CONICET, PIP 6160.

\section*{Appendix}
\subsection*{Conventions}
Noncommutativity affects the two spatial coordinates $x_i$
($i,j=1,2$),
$$[x_0,x_i]=0, [x_i,x_j]=i\theta\epsilon_{ij}\,,$$
and is realized in terms of the star product (\ref{eq:defmoyal}).

The Moyal-Weyl correspondence \cite{Sz},\cite{H} maps integration on NC space to
traces on Fock space as
\be
 \int d^2x\,f(x)=2\pi\theta\,\mathrm{Tr} [{\cal O}_f]
\ee The association between functions $f(\mathbf x)$ in NC space and
Weyl ordered operators ${\cal O}_f$ is via
\be
 {\cal O}_f(\hat {\mathbf x})=\int d^2x\, f(\mathbf x)\, \hat\Delta(\mathbf x)
\ee
where
\be
\hat\Delta(\mathbf x)=\int\frac { d^2k} {(2\pi)^2}\,e^{i\mathbf k\cdot(\hat {\mathbf x}-\mathbf x)}
\ee
One can see that derivatives on NC space can be implemented as
\be
 \partial_i f=\frac i \theta\epsilon_{ij}[x_j,f].
\ee
The matrix base functions ${b^{(i,j)}(\bf \mathbf x)}$ appearing in the text are the Weyl ordered
representation in NC space of the Fock space operators $|i\rangle\langle j|$. One
therefore has $[{b^{(i,j)}(\bf \mathbf x)}]^\dagger={b^{(j,i)}(\bf \mathbf x)}$.

\subsection*{On finite temperature  energy sums}

To separate the temperature dependence from the zero temperature contribution
in expressions (\ref{s}) and (\ref{t}) we used the identity
\be
\sum_{n=-\infty}^{\infty}\frac1{n^2+a^2}=\frac{2\pi}a\big[\frac12+\frac1{e^{2\pi a}-1}\big]
\ee

\newpage

\subsection*{Diagrammatics}

To keep track of the matrix indices of the field in (\ref{md}) a
double line notation is useful. The free theory two point propagator
(\ref{propagator}) can be drawn as
\begin{center}
\hspace{-1.2cm} \psset{xunit=1cm} \psline[linearc=.1]{>->}(0,0)(2,0)
\rput[bI](-.3,0.05){$i_1$} \rput[bI](2.25,.05){$j_2$}
\rput[bI](-.8,-.15){$_{-n}$} \psline{<-<}(0,-.15)(2,-.15)
\rput[bI](-.3,-.4){$i_2$} \rput[bI](2.25,-.4){$j_1$}
\rput[bI](2.7,-.15){$_{n}$}
\end{center}

~

\noindent The fourth order vertex (\ref{eq:defsibos}) in matrix base (\ref{fi}) is represented as

~

~
\begin{center}
\hspace{-2cm} \psset{xunit=1cm}
\psline[linearc=.1]{>->}(0,0)(.75,0)(.75,.75)
\rput[bI](-.3,0.05){$k_2$} \rput[bI](.65,.9){$k_2$}
\rput[bI](-.7,-.15){$_{n_1}$}
\psline[linearc=.1]{>->}(.90,.75)(.90,0)(1.65,0)(1.65,.75)
\rput[bI](1.05,.9){$k_3$} \rput[bI](.85,1.4){$_{n_2}$}
\rput[bI](1.55,.90){$k_3$}
\psline[linearc=.1]{>->}(1.80,.75)(1.80,0)(2.55,0)
\rput[bI](1.95,.9){$k_4$} \rput[bI](1.75,1.4){$_{n_3}$}
\rput[bI](2.8,.05){$k_4$}
\rput[bI](3.25,-.2){$_{n_4}$}
\psline{<-<}(0,-.15)(2.55,-.15)
\rput[bI](-.3,-.4){$k_1$}
\rput[bI](2.8,-.4){$k_1$}
\end{center}

\noindent The first order corrections to the free energy
(\ref{f1corr}) are
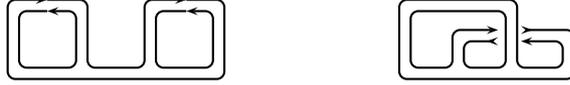
\vspace{.5cm}
\begin{center}
\hspace{-6.5cm}\psset{xunit=1cm}
\psline[linearc=.1]{<-}(.37,.75)(.75,.75)(.75,0)(0,0)(0,.75)(.37,.75)
\psline[linearc=.1]{->}(.37,.90)(.90,.90)(.90,0)(1.65,0)(1.65,.90)(2.2,.90)
\psline[linearc=.1]{->}(2.2,.75)(1.80,.75)(1.80,0)(2.55,0)(2.55,.75)(2.2,.75)
\psline[linearc=.1]{<-}(.37,.90)(-.15,.90)(-.15,-.15)(2.70,-.15)(2.7,.9)(2.2,.9)
\psline[linearc=.1]{<-<}(6.6,.35)(7.15,.35)(7.15,0)(6.55,0)(6.55,.90)(5,.90)(5,-.15)(7.3,-.15)(7.3,.5)
(6.6,.5)
\psline[linearc=.1]{>->}(6.3,.35)(5.85,.35)(5.85,0)(6.4,0)(6.4,.75)(5.15,.75)(5.15,0)(5.7,0)(5.7,.5)(6.3,.5)
\end{center}

\subsection*{Propagator corrections and self-energy}

The two point propagator function is defined as
\be
 D({\bf x}_1,t_1|\,{\bf x}_2,t_2)=\langle\phi({\bf x}_1,t_1)\,\phi({\bf x}_2,t_2)\rangle~.
\ee
Translational invariance in $t$ coordinate implies that $D$ is a
function of $t_1-t_2$. The kernel $\tilde D_n$ in Fourier space is
defined by (see (\ref{exp}))
\be
 D({\bf x}_1,t_1|\,{\bf x}_2,t_2)=\frac1\beta\sum_{n=-\infty}^\infty\sum_{i,j=0}^\infty\,e^{i\omega_n (t_1-t_2)}
 b^{(i_1,j_1)}({\bf x}_1)\, b^{(i_2,j_2)}({\bf x}_2)\; \tilde D_n^{(i_1,j_1;i_2,j_2)}
\ee
where $\omega_n$ are the Matsubara frequencies, and takes the form
\be
 \tilde D_n^{(i_1,j_1;i_2,j_2)}=\langle \phi_{-n}^{(i_1,j_1)}\phi_n^{(i_2,j_2)}\rangle~.
\ee
The first order corrections to the propagator in a diagrammatic expansion are

~

\begin{center}
\hspace{-12cm} \psset{xunit=1cm}
\psline[linewidth=2pt,linearc=.1]{>->}(0,0)(1,0)
\rput[bI](-.3,0.05){$i_1$} \rput[bI](1.25,.05){$j_2$}
\rput[bI](-.8,-.15){$_{-n}$}
\psline[linewidth=2pt]{<-<}(0,-.15)(1,-.15)
\rput[bI](-.3,-.4){$j_1$} \rput[bI](1.25,-.4){$i_2$}
\rput[bI](1.6,-.15){$_{n}$} \rput[bI](2,-.15){~$=$}
\psline{>->}(3.5,0)(4.5,0) \rput[bI](3.2,0.05){$i_1$}
\rput[bI](4.75,.05){$j_2$} \rput[bI](2.7,-.15){$_{-n}$}
\psline{<-<}(3.5,-.15)(4.5,-.15) \rput[bI](3.2,-.4){$j_1$}
\rput[bI](4.75,-.4){$i_2$} \rput[bI](5.1,-.15){$_{n}$}
\rput[bI](5.5,-.15){$+$} \hspace{.5cm}
\psline[linearc=.1]{>->}(6,0)(6.5,0)(6.5,.75)(7.5,.75)(7.5,0)(8,0)
\psline[linearc=.1]{<-}(6.65,.37)(6.65,.6)(7.35,.6)(7.35,0)(6.65,0)(6.65,.35)
\psline{<-<}(6,-.15)(8,-.150) \rput[bI](5.7,0.05){$i_1$}
\rput[bI](5.7,-.4){$j_1$} \rput[bI](5.3,-.15){$_{-n}$}
\rput[bI](6.9,.2){$k$} \rput[bI](8.3,0.05){$j_2$}
\rput[bI](8.3,-.4){$i_2$} \rput[bI](8.7,-.15){$_{n}$}
\rput[bI](9.1,-.15){$+$}\hspace{.5cm}
\psline{>->}(9.5,0)(11.5,0)
\psline[linearc=.1]{<-<}(9.5,-.15)(10,-.15)(10,-.9)(11,-.9)(11,-.15)(11.5,-.15)
\psline[linearc=.1]{<-}(10.15,-.50)(10.15,-.15)(10.85,-.15)
(10.85,-.75)(10.15,-.75)(10.15,-.52)
\rput[bI](9.3,0.05){$i_1$}
\rput[bI](9.3,-.4){$j_1$} \rput[bI](8.9,-.15){$_{-n}$}
\rput[bI](11.8,0.05){$j_2$}
\rput[bI](11.8,-.4){$i_2$} \rput[bI](12.2,-.15){$_{n}$}
\rput[bI](10.40,-.55){$k$}
\end{center}

~

\hspace{3.5cm}
\rput[bI](-1.5,-.30){$+$}
\psline[linearc=.1]{>->}(0,-.150)(.75,-.150)(.75,.75)(2.15,.75)(2.15,-.3)(0,-.3)
\psline[linearc=.1]{<-<}(1.95,.23)(1.22,.23)(1.22,-.15)(.9,-.15)(.9,.6)
(2,.6)(2,-.15)(1.37,-.15)(1.37,.08)(1.95,.08)
\psline{>->}(2.20,.23)(3,.23)
\psline{<-<}(2.20,.08)(3,.08)
\rput[bI](-.3,-.1){$i_1$}
\rput[bI](-.3,-.50){$j_1$}
\rput[bI](-.8,-.3){$_{-n}$}
\rput[bI](3.3,.28){$j_2$}
\rput[bI](3.3,-.2){$i_2$}
\rput[bI](3.6,0.05){$_n$}

\end{document}